# A few points on point-to-point correlation functions

Derek B. Leinweber[a]

[a]Department of Physics, The Ohio State University, 174 West 18th Avenue, Columbus, OH 43210-1106

The short-time regime of QCD two-point correlation functions is examined through a QCD-Sum-Rule-inspired continuum model. QCD Sum Rule techniques are tested and alternate nucleon interpolating fields are discussed. The techniques presented here may be of practical use in determining heavy-light meson decay constants.

## 1. INTRODUCTION

In the quest for an *ab initio* determination of the low-lying hadron spectrum, lattice QCD investigations have focused on the large Euclidean-time tails of three-momentum-projected two-point correlation functions. As such, the short time regime, where excited state contributions are significant, has simply been discarded.

The lattice approach to QCD allows the determination of correlation functions deep in the nonperturbative regime. However, the QCD Sum Rule (QCDSR) method [1] is restricted to the near perturbative regime of the Operator Product Expansion (OPE). In this regime, one cannot ignore the contributions of excited states in QCD correlation functions. To account for these contributions, the so-called "crude continuum model" is introduced [1]. The contributions of this model relative to the ground state, whose properties one is really trying to determine, are not small. The validity of this model is relied upon to effectively remove the excited state contaminations.

This investigation examines the physics in the near perturbative regime of point-to-point correlation functions where QCDSR analyses are performed. The QCD Continuum Model is constructed for three-momentum-projected Euclidean-time two-point functions following the techniques established in the QCDSR approach. This model will then be used as a probe of the physics represented in lattice QCD correlation functions, and as a test of QCDSR techniques.

Some attention has recently been given to the behavior of lattice point-to-point correlation functions in the near perturbative regime. There [2] the focus is on space-like separated correlation functions. It was concluded that the QCDSR inspired continuum model was sufficient to describe the lattice correlation functions over the calculated range. Using the entire lattice correlation function (including the deep nonperturbative regime), ground state masses were extracted and were found to agree with conventional lattice analyses.

Here the emphasis is on determining ground state properties by examining only the first few points of the lattice correlation function. This is similar in spirit to QCDSR analyses. In this manner, the validity of the continuum model is rigorously tested. By extending the analysis interval of the correlation function deeper into the nonperturbative regime, the evolution of fitted ground state properties may be monitored. Conventional lattice results are recovered when the interval extends deep into the nonperturbative regime. A sensitivity to the analysis interval in the extracted parameters would indicate a failure of the QCDSR inspired continuum model. This approach also allows an examination of the importance of the higher order terms of the OPE in constructing continuum models. These terms were not investigated in [2].

One would also like to evaluate whether these techniques are useful in analyzing lattice QCD correlation functions. Such techniques may be of practical use for analyzing two-point correlation functions which become noisy prior to a clear ground state domination, such as heavy-light meson correlators.

Finally these techniques will be exploited to investigate nucleon properties using unconventional interpolating fields. While there are formal field theoretic arguments indicating nucleon properties are independent of the interpolating field, it remains to demonstrate that this is in fact the case in practice.

## 2. THE CONTINUUM MODEL

Consider the following two-point function for the nucleon

$$G(t,\vec{p}) = \sum_{\vec{x}} e^{-i\vec{p}\cdot\vec{x}} \text{tr}\left[ \Gamma_4 \left\langle 0 \left| T\{\chi_N(x)\overline{\chi}_N(0)\} \right| 0 \right\rangle \right] \quad (1)$$

where

$$\chi_N(x) = \epsilon^{abc} \left( u^{Ta}(x) C\gamma_5 d^b(x) \right) u^c(x), \quad (2)$$

is the standard lattice interpolating field for the nucleon, and $\Gamma_4 = (1+\gamma_4)/4$ projects positive parity states for $\vec{p} = 0$. At the phenomenological level, one inserts a complete set of states $N_i$ and defines

$$\left\langle 0 \left| \chi_N(0) \right| N_i, p, s \right\rangle = \lambda_{N_i} u(p,s), \quad (3)$$

where the coupling strength, $\lambda_{N_i}$, measures the ability of the interpolating field $\chi_N$ to annihilate the i'th nucleon excitation. For $\vec{p} = 0$ and Euclidean time $t \to \infty$, the ground state dominates and

$$G(t) \to \lambda_N^2 e^{-M_N t}. \quad (4)$$

The spectral representation is defined by

$$G(t) = \int_0^\infty \rho(s) e^{-st} ds, \quad (5)$$

and the spectral density is $\rho(s) = \lambda_N^2 \delta(s-M_N) + \zeta(s)$ where $\zeta(s)$ provides the excited state contributions.

While it may be tempting to fit the correlation function in the shorter time regime by including additional poles in the spectral density, such an approach fails in a number of ways [3]. The correlation function is probably best described by many states of diminishing coupling strengths and increasing widths. It may be that the QCDSR inspired continuum model is an efficient way of characterizing this physics.

The form of the spectral density used in the continuum model is determined by the leading terms of the OPE surviving in the limit $t \to 0$. In Euclidean space and coordinate gauge the quark propagator has the expansion

$$\begin{aligned} S_q^{aa'} &= \frac{1}{2\pi^2} \frac{\gamma \cdot x}{x^4} \delta^{aa'} + \frac{1}{(2\pi)^2} \frac{m_q}{x^2} \delta^{aa'} \\ &\quad - \frac{1}{2^2 3} \left\langle :\overline{q}q: \right\rangle \delta^{aa'} + \cdots, \end{aligned} \quad (6)$$

and $G(t)$ has the following OPE

$$\begin{aligned} G(t) &\simeq \frac{3 \cdot 5^2}{2^8 \pi^4} \left( \frac{1}{t^6} + \frac{28}{25} \frac{m_q a}{t^5} + \frac{14}{25} \frac{m_q^2 a^2}{t^4} \right. \\ &\quad \left. - \frac{56\pi^2}{75} \frac{\left\langle :\overline{q}q: \right\rangle a^3}{t^3} + \cdots \right). \end{aligned} \quad (7)$$

The spectral density used in the continuum model is defined by equating (5) and (7). The continuum model is defined through the introduction of a threshold which marks the effective onset of excited states in the spectral density.

$$\int_{s_0}^\infty \rho(s) e^{-st} ds \quad (8)$$

$$= e^{-s_0 t} \sum_{n=1}^6 \sum_{k=0}^{n-1} \frac{1}{k!} \frac{s_0^k}{t^{n-k}} C_n \mathcal{O}_n,$$

where $C_n$ and $\mathcal{O}_n$ are the Wilson coefficient and normal ordered operator of the term $t^{-n}$ in (7). The phenomenology of $G(t)$ is summarized by

$$G(t) = \lambda_N^2 e^{-M_N t} + \xi \int_{s_0}^\infty \rho(s) e^{-st} ds. \quad (9)$$

Strictly speaking, $\xi = 1$ but here is optimized with $\lambda_N$, $M_N$, and $s_0$ to account for enhancement of the correlator in the short time regime due to lattice anisotropy. Ref. [2] found the anisotropy to be large for $x - x_0 < 6$ for free quark correlators and remain large in their interacting simulation at $\beta = 5.7$. At $\beta = 5.9$ there is some hope that anisotropy issues will be less problematic for the Fourier transformed correlators presented here. However, at very short times the quarks are essentially free and the anisotropy must be accommodated.

The nucleon coupling strength, $\lambda_N$, is determined in absolute terms, without resorting to a ratio of the continuum contributions as done in [2]. Our approach requires the use of tadpole-improvement [4]

$$\sqrt{2\kappa} \to \left(1 - \frac{3\kappa}{4\kappa_{cr}}\right)^{1/2} \qquad (10)$$

to account for otherwise large renormalization factors. Remaining renormalization associated with composite operators is argued to be small [3]. The $\kappa$ dependence of these two wave function normalizations is very different, and is crucial to recovering the correct mass independence of the Wilson coefficient of the identity operator.

With this approach, the effects of lattice anisotropy may be absorbed through a combination of a larger continuum strength ($\xi > 1$) and marginally larger continuum threshold ($s_0$). To test the QCDSR method as closely as possible to its actual implementation, we fix the value for $\xi$ obtained in a four parameter fit from $t = 6 \to 20$. This value is similar to that obtained by fitting (7) to the first few time slices of the lattice data. The value of $\xi$ depends on the choice of nucleon interpolating field and is largest for $\chi_N$ of (2) at 10.0(3). This suggests the presence of significant anisotropy at the second time-slice following the source [3].

Infrared lattice artifacts are not a significant problem for this approach. The ultraviolet lattice cutoff is modeled in a manner similar to that for the continuum model. By the second time slice following the source, such corrections are found to be negligible. In the following we simply discard the source ($t = 4$) and first ($t = 5$) time slices from the fit of the correlation functions.

## 3. NUCLEON CORRELATOR FITS

Our purpose is to test whether the nucleon mass and coupling strength can be obtained accurately from a fit considering only the first few points of the correlation function. The lattice correlation functions [5] are fit with (9) in a three parameter search of $\lambda_N$, $M_N$ and $s_0$, in analysis intervals from $t = 6 \to t_f$ where $t_f$ ranges from 9 through 23. The nucleon mass determined in each of the intervals is plotted as a function of $t_f$ in Figure 1. $s_0$ displays a similar plateau.

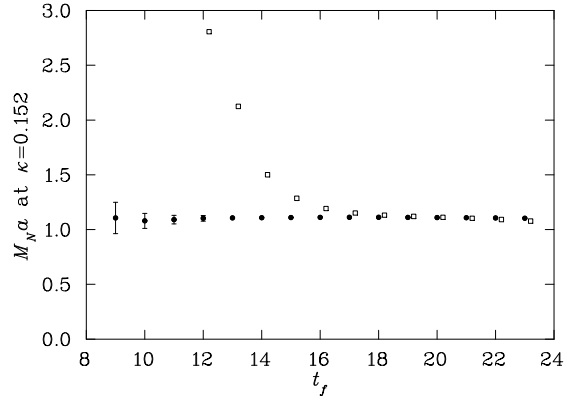

Figure 1. The nucleon mass determined in each analysis interval plotted as a function of $t_f$. In this and the following figure, bullets correspond to pole plus continuum fits from $t = 6 \to t_f$, and open squares illustrate a simple pole fit to the region $t_f - 7 \to t_f$, which is selected to give similar uncertainties in the nucleon mass at $t = 20$.

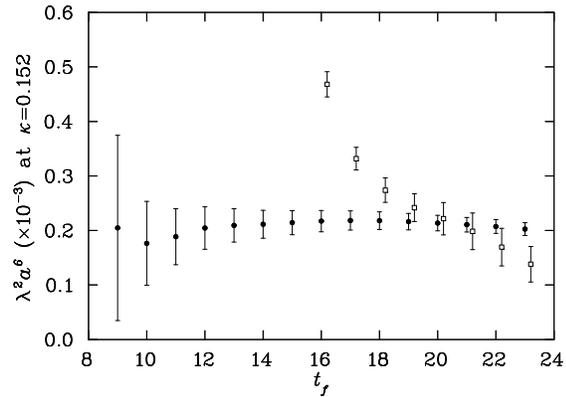

Figure 2. The nucleon coupling strength determined in each analysis interval plotted as a function of $t_f$. Symbols are as in Figure 1.

Figure 2 illustrates similar results for the coupling strength. It is interesting to see that the simple pole determination of $\lambda_N$ fails to form a plateau at large time separations. Similar results are seen for the larger values of $\kappa = 0.154$ and 0.156. The plateau in $M_N$ and $\lambda_N$ for pole plus continuum fits indicates that the continuum model effectively accounts for excited state con-

Table 1
Summary of Lattice Results for OPE-$m_q$-improved continuum model.

| Interpolating Fields | $M_N$ (GeV) | $\lambda_N$ (GeV$^3$) | $\xi$ | $s_0$ (GeV) |
|---|---|---|---|---|
| $\chi_1 = \epsilon^{abc}\left(u^a\, C\gamma_5\, d^b\right) u^c$ | 0.938* | 0.016(3) | 10.0(3) | 1.97(12) |
| $\chi_2 = \epsilon^{abc}\left(u^a\, C\, d^b\right)\gamma_5 u^c$ | Not seen | 0.0002(3)† | 2.43(12) | 1.84(9) |
| $(\chi_1\overline{\chi}_2 + \chi_2\overline{\chi}_1)/2$ | Fixed | 0.0018(14) | 2.46(15) | 1.76(22) |
| $\chi_{\rm SR} = \epsilon^{abc}\left(u^a\, C\gamma_\mu\, u^b\right)\gamma_5\gamma^\mu d^c$ | 0.96(4) | 0.035(6) | 6.6(2) | 2.18(12) |
| $\chi_\Delta = \epsilon^{abc}\left(u^a\, C\gamma^\mu\, u^b\right) u^c$ | 1.13(4) | 0.027(7) | 10.7(3) | 2.39(10) |

*Defines the lattice spacing $a$.      †Inferred from $\chi_1\overline{\chi}_1$ and $\chi_1\overline{\chi}_2$ results.

taminations in the correlation functions. However, interplay between the fit parameters give rise to rather large uncertainties at $t_f = 9$.

## 4. ALTERNATE INTERPOLATORS

These techniques are now used to investigate nucleon properties obtained from correlation functions of unconventional nucleon interpolating fields which become noisy prior to a clear ground state domination. The results are summarized in Table 1.

There is a long history of argument over the optimum nucleon interpolating field to be used for QCDSR analyses. The solution is now clear. This analysis indicates that the continuum model effectively removes excited state contaminations and allows the isolation of the ground state. Hence, one should choose interpolating fields such as $\chi_{\rm SR}$ [6] that give good convergence of the OPE, as the corresponding increase in continuum contributions is tolerable.

Ground state nucleon properties *are* independent of the interpolating field used to excite the baryon from the vacuum. This invariance is satisfied in a trivial manner. The interpolating field $\chi_2$, which vanishes in the nonrelativistic limit, has negligible overlap with the nucleon ground state.

The results of Table 2 for $\lambda_N$, evolved to a scale of 1 GeV$^2$ in the leading log approximation, compare favorably with other approaches. $s_0$ is somewhat large as anticipated above. The smaller value of $\lambda_N$ for Ref. [2] may be due to the omission of large $\alpha_s$ corrections to the Wilson coefficient of the identity operator used to normalize $\lambda_N$. Such corrections could increase their result by 20%.

Table 2
Comparison with selected results for $\chi_{\rm SR}\overline{\chi}_{\rm SR}$.

| Approach | $\lambda_N$(1 GeV$^2$) (GeV$^3$) | $s_0$ (GeV) |
|---|---|---|
| This work | 0.030(6) | 2.18(12) |
| Ref. [7] | 0.032(1) | 1.92(5) |
| Ref. [8] | 0.031(6) | 1.69(15) |
| Ref. [2] | 0.022(4) | <1.4 |

The correlation functions used in this analysis were obtained in collaboration with Terry Draper and Richard Woloshyn [5]. This research is supported in part by the Department of Energy and the National Science Foundation.